\documentclass[aps,prl,reprint,amsmath, amssymb, aps,superscriptaddress,nolongbibliography]{revtex4-2}
\usepackage{bm}
\usepackage{color}
\usepackage{graphicx}
\usepackage{comment}
\usepackage{makecell}

\definecolor{DarkGreen}{rgb}{0.0, 0.5, 0.0}
\definecolor{CorrectionPurple}{rgb}{0.65,0.34,0.65}
\usepackage[colorlinks,citecolor=DarkGreen,linkcolor=blue, urlcolor=blue,linktocpage]{hyperref}

\newcommand{\so}[1]{{\color{red}{#1}}}
\renewcommand{\so}[1]{#1}
\newcommand{\coauthor}[1]{{\color{red}{#1}}}
\newcommand{\coauthormatsu}[1]{{\color{red}{#1}}}
\renewcommand{\coauthor}[1]{#1}
\renewcommand{\coauthormatsu}[1]{#1}

\newcommand{\soo}[1]{{\color{red}{#1}}}
\newcommand{\mo}[1]{{\color{red}{#1}}}

\renewcommand{\soo}[1]{#1}
\renewcommand{\mo}[1]{#1}

\begin{document}
\title{Orbital Paramagnetism without Density of States Enhancement \\ in Nodal-line Semimetal ZrSiS}

\author{Soshun Ozaki}
\affiliation{Department of Basic Science, University of Tokyo, Meguro, Tokyo 153-0041, Japan}
\author{Hiroyasu Matsuura}
\affiliation{Department of Physics, University of Tokyo, Bunkyo, Tokyo 113-0033, Japan}
\author{Ikuma Tateishi}
\affiliation{Department of Physics, Osaka University, Toyonaka, Osaka 560-0043, Japan}
\author{Takashi Koretsune}
\affiliation{Department of Physics, Tohoku University, Sendai, Miyagi 980-8578, Japan}
\author{Masao Ogata}
\affiliation{Department of Physics, University of Tokyo, Bunkyo, Tokyo 113-0033, Japan}
\affiliation{Trans-Scale Quantum Science Institute, University of Tokyo, Bunkyo, Tokyo 113-0033, Japan}
 %\\

\begin{abstract}
%%%%%%%%%%%%%%%%%%%%%%%%%%%
% Abstract %%%%%%%%%%%%%%%%
%%%%%%%%%%%%%%%%%%%%%%%%%%%
\begin{comment}
%Orbital paramagnetism is typically associated with a large density of states (DOS). 
%However, recent findings in the nodal-line semimetal ZrSiS reveal anomalous orbital 
%paramagnetism without DOS enhancement, suggesting an unconventional mechanism.
We propose a mechanism of the orbital paramagnetism newly observed in the nodal-line semimetal ZrSiS, which occurs without the usual density of states (DOS) enhancement.
Based on the density functional theory calculations and effective model analysis, 
%we elucidate the anomalies 
%in the orbital magnetic susceptibility:
%the unconventional orbital paramagnetism, its temperature dependence, 
%and its anisotropy.
we show that the fluctuation of the energy at the node points along the nodal line,
inherent in the realistic nodal-line systems,
leads to the orbital paramagnetism without enhanced DOS.
This orbital paramangetism is understood in terms of the interband effect between bands with negative curvature.
Our result indicates that the observed paramagnetism in ZrSiS is clear evidence for 
the existence of the predicted nodal line.
%Our results indicate that the observed anomalous orbital paramagnetism is due to
%the interband effect originating from the fluctuation of energy along the nodal line, 
%inherent in the realistic nodal-line systems.
%The orbital paramagnetism is understood based on 
%a simple two-band effective model with negative curvature.
%(571 letters)
\end{comment}
%240604 version
Unconventional orbital paramagnetism without enhanced density of states was recently discovered in the nodal-line semimetal ZrSiS. We propose a novel interband mechanism, linked to the negative curvature of energy dispersions, which successfully accounts for the observed anomalous response. This negative curvature originates from energy variation along the nodal line, inherent in realistic nodal-line materials. Our results suggest that such orbital paramagnetism provides strong evidence for the presence of nodal lines in ZrSiS, and serves as a hallmark of other nodal-line materials.
\end{abstract}

%%% Keywords are not needed any longer. %%%
%%%\kword{keyword1, keyword2, keyword3, \ldots}
%%%

\date{\today}
\maketitle
%%%%%%%%%%%%%%%%%%%%%%%%%%%%%%%%%%%%%%%%%%%%%%%%%%%%%
% Introduction %%%%%%%%%%%%%%%%%%%%%%%%%%%%%%%%%%%%%%
%%%%%%%%%%%%%%%%%%%%%%%%%%%%%%%%%%%%%%%%%%%%%%%%%%%%%
%\textit{Introduction}.--- 
Orbital magnetism is one of the fundamental properties of solids, rooted in the seminal work of Landau and Peierls on free-electron and tight-binding models \cite{landau,peierls}. 
In Dirac electron systems such as bismuth and graphene, significant orbital diamagnetism arises from the interband effect of a magnetic field \cite{wherli,fukuyama-kubo,fukuyama2007,koshino-ando,fujiyama}. 
This discovery has shown that orbital magnetism is highly sensitive to the electronic structure, leading to extensive ongoing research into their relationship.
\coauthor{Orbital magnetism is a bulk response that is also sensitive to the geometric properties encoded in the wave functions, such as nontrivial topology and quantum geometry. 
Weyl semimetals \cite{wherli,fujiyama,mcclure,fukuyama-kubo,fukuyama2007,nakamuradirac, nakamuradirac,sharapov,ghosal,koshino-ando,ogata3, kariyado}, topological insulators
\cite{murakami,nakai-nomura,ozakiogata,ozakiogatalong},
and nodal-line semimetals \cite{burkov2011topological,fang2015topological,tateishi2018face,tateishi2020mapping,tateishi2020nodal,koshino,tateishi, fujiyama,ozakihmtsf}
are prominent examples of these connections.
Among these, nodal-line semimetals have been less characterized
experimentally. Consequently, orbital magnetism is expected to offer a more
direct, bulk-sensitive probe reflecting their characteristic electronic structures, complementing standard
techniques such as angle-resolved photoemission spectroscopy \cite{neupane2016observation,bian2016drumhead, chan20163}, quantum oscillations \cite{hu2016evidence,mikitik_zrsis,gudac}, and transport studies \cite{endo, mizoguchi, hosoi,ogataozaki}.}

%Recently, unconventional orbital responses have been observed 
%in the nodal-line semimetal ZrSiS at low temperatures, when a magnetic field is applied along the $C_4$ rotation axis \cite{gudac}. 
%This behavior is not explained by the Pauli paramagnetism estimated from the observed low density of states (DOS). 
\coauthor{Orbital \textit{paramagnetism}, as an unconventional orbital response, has recently been observed in the 
nodal-line semimetal ZrSiS at low temperatures, when a magnetic field is applied along the $C_4$ rotation axis \cite{gudac}. 
This finding challenges the established understanding based on Larmor's classical picture and Landau's theory, both of which regard orbital magnetism as 
inherently diamagnetic.}
Furthermore, as the temperature increases, the paramagnetism 
decreases and eventually changes to diamagnetism at around 120~K, which is also an unexpected behavior. 
A similar 
paramagnetism has also been observed in another nodal-line semimetal, SrAs$_3$ \cite{Hosen2020}. 
\coauthor{These observations in nodal-line semimetals suggest a possible 
connection between the presence of nodal lines and the emergence of 
such orbital paramagnetism, % \cite{gudac,mikitik}, 
though its microscopic origin remains elusive.}
%Despite several efforts to 
%understand these anomalous behaviors, their microscopic origins, likely related to the nodal line \cite{gudac,mikitik}, remain elusive.

%\textcolor{red}{Theoretically, orbital magnetism has long been studied mainly as \textit{diamagnetism}.
%Contrary to physical intuition as exemplified by Larmor diamagnetism,}
At present, a few origins of orbital paramagnetism are known, including
the Van Hove singularity \cite{vignaleparasaddlepoint,raoux-piechon, bustamante2023}, flatband \cite{Rhim2020, piechon}, 
and other mechanisms \cite{mikitik,schober}. 
In most of these cases, however, 
orbital paramagnetism is accompanied by a strong enhancement of density of states (DOS), 
and thus it is often masked by the large Pauli paramagnetism. 
%\begin{comment}
In this Letter, we propose a new mechanism to explain the orbital paramagnetism without enhanced DOS.
We first present a quantitative analysis of the orbital magnetic susceptibility in ZrSiS using 
the effective models based on the density functional theory (DFT) calculations, 
successfully elucidating the observed paramagnetism, temperature dependence, and anisotropic behaviors \cite{gudac,shao2024}. 
To further understand the origin of this orbital paramagnetism, we derive a simple effective model. 
Our results show that the observed orbital paramagnetism without DOS enhancement is attributed to the interband effect associated with negative curvature in the energy dispersion. 
This negative curvature originates from the energy 
variation along the nodal line, which is inherent in realistic nodal-line materials. 
\coauthor{While dispersions with negative curvature can in principle occur in other systems, they emerge more naturally and systematically in nodal-line 
semimetals due to the presence of line-like band crossing.}
%Importantly, our findings unveil a novel mechanism for orbital paramagnetism, successfully explaining that without an enhancement of the DOS. 
This novel mechanism suggests that such orbital paramagnetism provides strong evidence for the nodal lines not only in ZrSiS but also potentially in various other nodal-line materials.

\textit{DFT calculations and models}.---
\begin{figure*}
\includegraphics[angle=0,width=\linewidth]{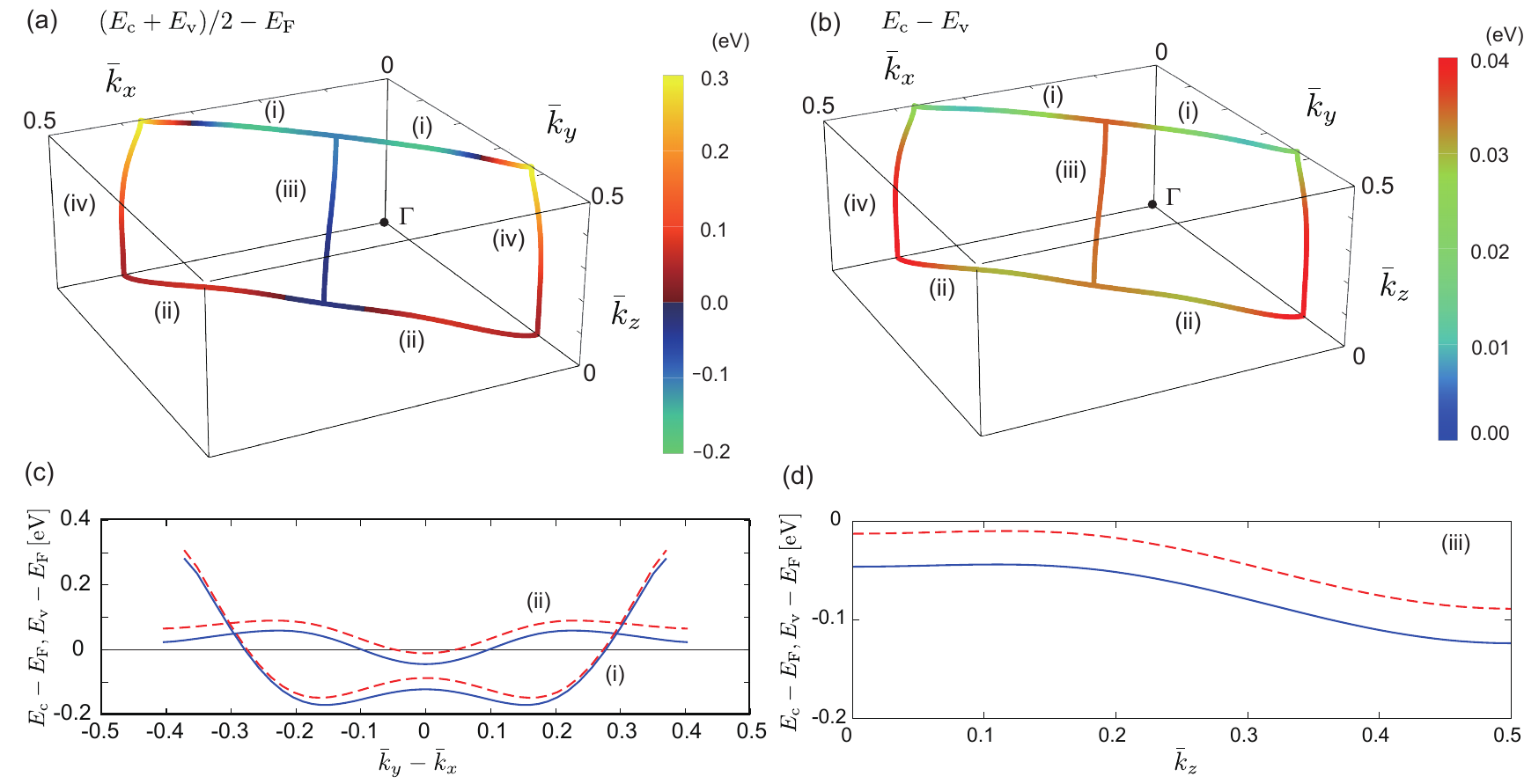}
	\caption{
	Positions of the nodal points in the Brillouin zone.
	The dimensionless wave numbers 
	$\bar{k}_x=k_x a/(2\pi)$, $\bar{k}_y=k_y a/(2\pi)$, and $\bar{k}_z=k_z c/(2\pi)$ are shown.
	(a) Color indicates the energy of the middle of the gap $(E_c({\bm k})+E_v({\bm k}))/2-E_F$ 
	with $E_c({\bm k})$ and $E_v({\bm k})$ being the energy of the bottom (top) of the conduction (valence) band
	at $\bm k$, and $E_{\rm F}$ is the Fermi energy.
	(b) Color indicates the magnitude of the gap due to the SOI, $E_c({\bm k})-E_v({\bm k})$.
	(c, d) Energy dispersions near $E_{\rm F}$ (c) along the nodal lines (i) and (ii), \soo{and (d) along the nodal line (iii)}. The valence (conduction) band is shown in solid (dashed) line.
	}
\label{fig:nlgap}
\end{figure*}
ZrSiS has a $P4/nmm$ structure \cite{haneveld,Tremel1987} and has a set of nodal lines \cite{schoop2016,gudac,rudenko,habe}.
To capture the characteristics of these nodal lines in detail,
we perform DFT calculations 
using \textsc{Quantum-ESPRESSO} and \textsc{Wannier90} packages \cite{Giannozzi2017,pbe1996,vanderbilt1990,Corso2014,Pizzi2020,koretsune2023}
with the lattice parameters of
$a=3.55$ \AA\ and $c=8.07$ \AA\ \cite{Onken1964}.
The details of these calculations are shown in Supplemental Material (SM) \cite{sm}.
Figure~\ref{fig:nlgap} shows the positions of the nodal points in the Brillouin zone,
at which the bottom of the conduction band ($E_c({\bm k})$) and the top of the valence band
($E_v(\bm k)$) are close to each other.
The colors in Fig.~\ref{fig:nlgap}(a) indicate the energy of the middle of the gap relative to the Fermi energy $E_F$,
i.e., $(E_c(\bm k)+E_v (\bm k))/2-E_F$, and those in Fig.~\ref{fig:nlgap} (b) indicate the magnitude of the gap $E_c(\bm k)-E_v(\bm k)$ due to the spin-orbit interaction (SOI). 
Four nodal lines (i) -- (iv) are indicated, which are in the planes of
$k_z=\pi/c$, $k_z =0$, $|k_x|=|k_y|$, and $k_x=0$ or $k_y=0$, respectively.
Note that the nodal line protected by the nonsymmorphic symmetry \cite{schoop2016} is omitted 
because it is approximately 1eV away from $E_F$.
From Fig.~\ref{fig:nlgap}(a), we can see that the nodal lines (i) and (ii) form closed lines in the Brillouin zone on the $k_zc=\pi$ and $k_zc=0$ planes, respectively, both of which cross $E_F$. 
As shown in Fig.~\ref{fig:nlgap} (c), the energy deviations around $E_F$ are approximately $\pm 0.25$ eV for (i) and $\pm 0.05$ eV for (ii), which play important roles in the later calculations. 
%The energy dispersions near $E_{\rm F}$ along the nodal lines (i) and (ii) are shown in Fig.~\ref{fig:nlgap}(c). 
\soo{Figure~\ref{fig:nlgap}(d) represents the energy of the valence and conduction bands along 
the nodal line (iii), where the Fermi level lies outside the Dirac mass gap.}
Nodal lines (iii) and (iv) form one-dimensional lines along the $k_z$ direction. 
While the nodal line (iii) is located close to $E_F$, (iv) is away from $E_F$, so that we neglect the nodal line (iv) in the following. 

\begin{figure}
\rotatebox{0}{\includegraphics[angle=0,width=\linewidth]{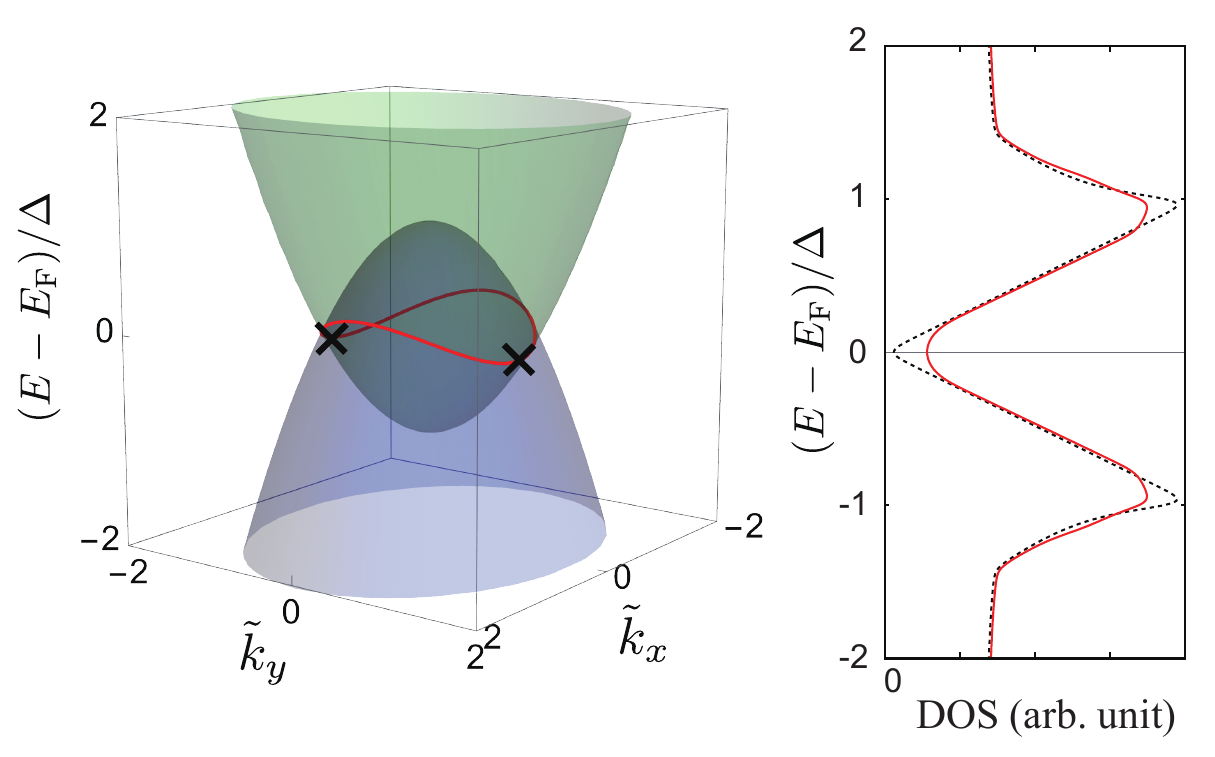}}
	\caption{(Left panel) Energy dispersions and DOS for model \eqref{eq:model} at $k_z=0$ 
	with $\tilde{\eta}=\eta k_{\rm R}^2/\Delta=0.2$. 
The red line indicates the nodal line. 
	$\tilde{k}_x=k_x/k_{\rm R}$ and $\tilde{k}_y=k_y/k_{\rm R}$ are dimensionless wave numbers. 
	The symbol $\boldsymbol \times$ indicates the nodal point at which the tangent of the nodal line is 
	parallel to $\tilde{k}_x$.
    \so{(Right panel) Density of states for $\tilde{\eta}=0$ (black dashed line) and $\tilde{\eta}=0.2$ (red solid line).}}
\label{fig:energyshifts}
\end{figure}

Considering these characteristics of ZrSiS, we introduce an effective model for the nodal lines (i) and (ii) as
\begin{equation}
	{\mathcal H}^\text{(X)} = \biggl[\frac{\hbar^2}{2m^*}(k_x^2 + k_y^2)-\Delta\biggr] \sigma_{z} + \hbar v_z k_z \sigma_{x} + \eta^\text{(X)} (k_x^2 - k_y^2) \sigma_0,
  \label{eq:model}
\end{equation}
%to understand the temperature dependence of magnetic susceptibility $\chi$.
%The effect of the nodal line (iii) will be considered later.
%In Eq.~\eqref{eq:model}, 
where X=i or ii is the index for the nodal lines and $\sigma_0$, $\sigma_{x}$, and $\sigma_{z}$ are the identity and Pauli matrices.
\so{We have assumed a circular nodal line instead of the expected square nodal line %We do not expect this difference to have a substantial impact on the result since they are topologically equivalent.}
}
\coauthor{for simplicity. This will be justified later by a comparison with the
experimental data.}
The first term represents the two parabolic bands with masses of different signs. 
As shown in Fig.~\ref{fig:energyshifts}, the overlap of the bands is $2\Delta$.
The second term in Eq.~\eqref{eq:model} hybridizes the two bands where $v_z$ represents the velocity in the $z$ direction. 
The ${\bm k}$ points on the nodal line satisfy $\frac{\hbar^2}{2m^*} (k_x^2+k_y^2)-\Delta=0$ and $k_z=0$,
indicating that the nodal line forms a circle in the $k_x$-$k_y$ planes at $E_F=0$.
\so{Finite $k_z$ opens gaps along the nodal line.}
The last term introduces variation of energy along the nodal line, and the energy deviation is $\pm2\eta^\text{(X)} k_{\rm R}^2$
with $k_{\rm R}=\sqrt{2m^*\Delta}/\hbar$ being the radius of the nodal line.
We will see below that this variation play a crucial role 
for the emergence of orbital paramagnetism.
Similar energy variation has been discussed in Refs.~\cite{mikitik, mikitik2,kobayashi2008,Suzumura_2017, ozakihmtsf}.
In the following, we omit the superscript of $\eta$ when a distinction is not necessary.
\so{The DOS is also shown in Fig.~\ref{fig:energyshifts} for $\tilde{\eta}=0$ and $0.2$,
where $\tilde{\eta}$ is a dimensionless parameter $\tilde{\eta}=\eta k_{\rm R}^2/\Delta$.
The energy variation results in the emergence of DOS around $E=E_F$.}
\so{This variation also gives negative \mo{Gauss} curvature \mo{of the $E$-$k_x$-$k_y$ surface} at $(k_x,k_y,k_z)=(0,\pm k_{\rm R}, k_{z0})$ for the valence band and at $(\pm k_{\rm R}, 0, k_{z0})$ for the conduction band for $k_{z0}\neq0$, i.e., in the presence of gaps.
The curvature is illustrated in detail in SM \cite{sm}.}
%Figure~\ref{fig:energyshifts} shows the energy dispersion and the nodal line (shown in red) for
%finite $\eta$.
Note that the effective model of Eq.~\eqref{eq:model} does not %have the gap.
\coauthor{include the SOI effect.
As discussed later, SOI hardly affects our results.}

The nodal line (iii) is approximately regarded as 2D Dirac electrons, governed by 
\begin{equation}
\mathcal{H}^\text{(iii)}=\hbar v_{\rm Dirac}(k_x\sigma_x + k_y \sigma_y ) + \Delta_{\rm SOI}\sigma_z + \varepsilon_0(k_z), \label{eq:nodal3}
\end{equation}
where $\varepsilon_0(k_z)$ is the energy shift and 
$\Delta_{\rm SOI}$ represents the gap originating from SOI.
We discuss the orbital magnetic susceptibility using models 
\eqref{eq:model} and \eqref{eq:nodal3} in the following.

%%%%%%%%%%%%%%%%%%%%%%%%%%
\textit{Orbital magnetic susceptibility}.---
The orbital magnetic susceptibility in a magnetic field in the $z$ direction ($\chi_z$) is generally given by \cite{fukuyama}
\begin{eqnarray}
	\chi_z = \frac{e^2}{\hbar^2}k_BT \sum_{n{\bm k}}{\rm Tr}
\, \gamma_x{\mathcal G}\gamma_y{\mathcal G}
\gamma_x{\mathcal G}\gamma_y \mathcal{G}, \label{eq:fukuyama}
\end{eqnarray}
where ${\mathcal G}=\mathcal{G}({\bm k},i\varepsilon_{n})=[i\varepsilon_n -\mathcal{H}^\text{(X)}+\mu]^{-1}$ is the thermal Green's function,
$\varepsilon_n = (2n + 1)\pi k_{\rm B} T$ ($n\in \mathbb{Z}) $ is the Matsubara frequency, and 
$\gamma_i$ is the velocity operator of in the $i$ ($i=x,y,z$) direction defined by $\gamma_i = \frac{\partial {\mathcal H}^\text{(X)}}{\partial k_{i}}$.
\so{Equation~\eqref{eq:fukuyama} is a general formula for noninteracting systems and incorporates the spin degeneracy}.
The orbital magnetic susceptibilities in other directions, $\chi_x$ and $\chi_y$, are obtained by a cyclic replacement of $x$, $y$, and $z$.
We numerically evaluate Eq.~\eqref{eq:fukuyama} for the model Eq. \eqref{eq:model}, employing the quasi-Monte Carlo method \cite{qmc,conroy1967, cranley1976, korobov1957, korobov1963} for ${\bm k}$ summation, and the sparse-ir method 
\cite{shinaoka, sparsesample,wallerberger}
for Matsubara summation.
\soo{The momentum cutoffs are $\Lambda_R=1000k_R$ in the radial direction and $\Lambda_{z}=(l_z)^{-1}$ in the $k_z$ direction
with $l_z=\hbar v_z/\Delta$.
The Matsubara summation includes a sufficient number of frequencies corresponding to the energy cutoff $\varepsilon_{\rm cut}=100 \Delta$}
\footnote{In our approach, we adopt a finite but sufficiently large cutoff for Matsubara summation. Correspondingly, we choose the ${\bm k}$ integral cutoff to be much larger than the Matsubara frequency cutoff.
This method confirms that the numerical integration can be carried out efficiently and with sufficient accuracy, when compared with the analytical results obtained in Ref.~\cite{tateishi} and SM \cite{sm}}.

\begin{figure}
    \includegraphics[width=1\linewidth]{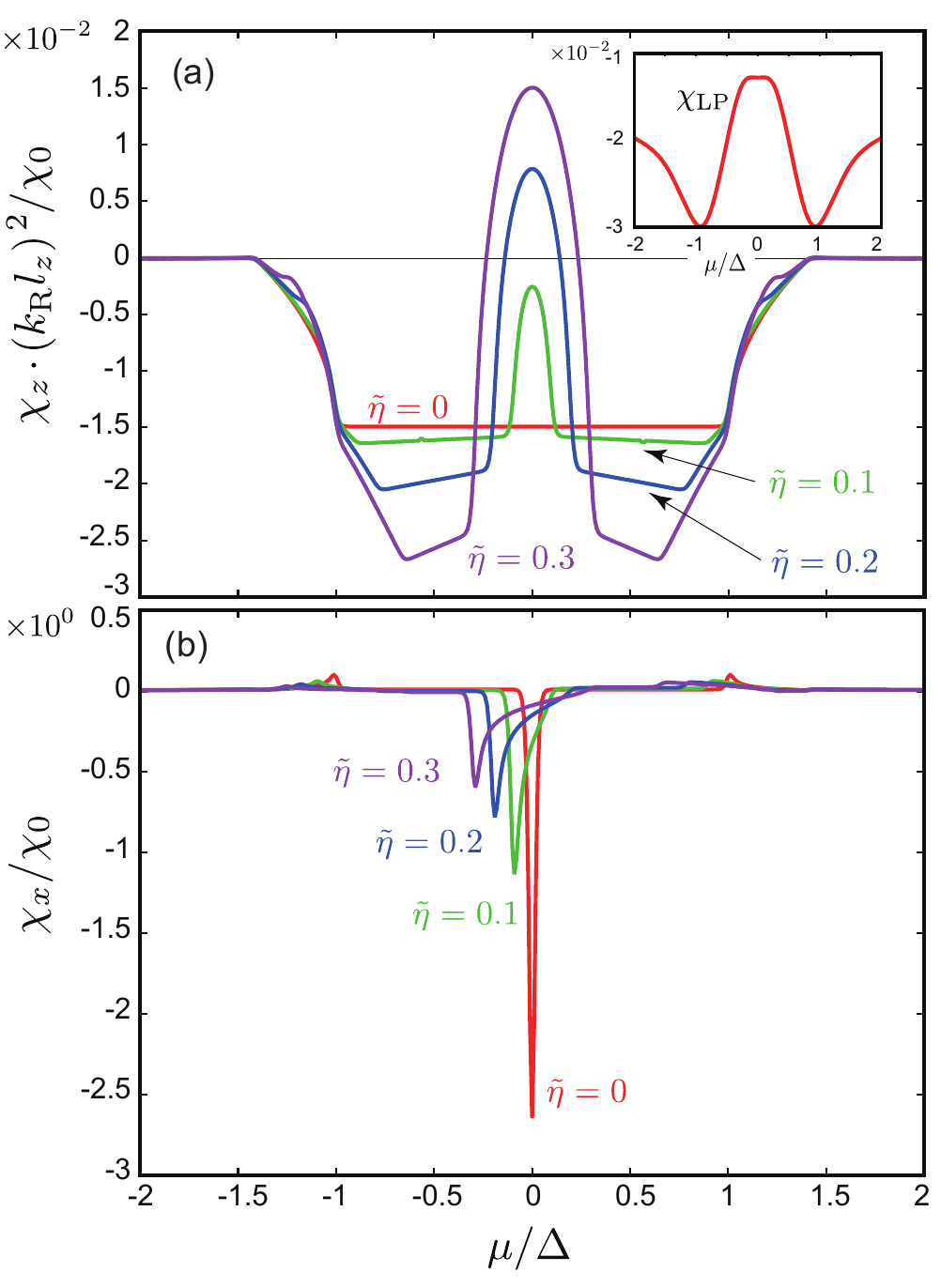}
	\caption{(a) Orbital magnetic susceptibility $\chi_z$ as a function of chemical potential
	when $k_BT=0.01\Delta$ for several values of $\tilde{\eta}=\eta k_{\rm R}^2/\Delta$.
	(b) Orbital magnetic susceptibility $\chi_x$ when $k_BT=0.05\Delta$. 
    They are normalized by $\chi_{0} \equiv \frac{e^2}{\hbar^2}Vl_z \Delta$ with 
	$V$ being the volume and $l_z=\hbar v_z/\Delta$.
    Inset: LP contribution for \soo{$\tilde{\eta}=0.2$} and $k_BT=0.1\Delta$.
	}
    \label{fig:chiz}
\end{figure}

The results for $\chi_z$ %at $k_BT=0.01\Delta$ 
and $\chi_x$ %at $k_BT=0.05 \Delta$ 
are shown in Fig.~\ref{fig:chiz} (a) and (b), respectively, for several values of 
$\tilde{\eta}$.
For the case without the energy variation ($\tilde{\eta}=0$), $\chi_z$ is diamagnetic for every $\mu$ 
while $\chi_x$ has a sharp peak at $\mu=0$ originating from the interband effect of the 2D Dirac electrons \cite{koshino-ando}.
We confirmed that our calculation reproduces the previous results \cite{tateishi} 
in the limit of $\Lambda_{\rm R}\to \infty$ and $\Lambda_z\to \infty$ [see SM\cite{sm} for detail].

Finite $\eta$ gives a significant effect on $\chi_z$: $\chi_z$ has a broad peak around $\mu=0$, whose width is approximately $2\eta k_{\rm R}^2$.
Furthermore, for $\tilde{\eta}>0.1$, the value at the peak is positive, meaning orbital paramagnetism.
The inset of Fig.~\ref{fig:chiz} (a) shows the Landau-Peierls (LP) contribution \cite{landau,peierls, raoux-piechon, ogata-fukuyama}, or the \textit{intraband}
contribution, which is negative for all values of $\mu$.
Therefore, we conclude that the obtained orbital paramagnetism near $\mu=0$ is due to an \textit{interband} effect.
As we noted before, while the orbital paramagnetism is usually accompanied by large DOS,
the present result suggests an \textit{interband orbital paramagnetism} without the 
enhancement of the DOS.
\so{The presence of two energetically close bands involving negative curvature is a crucial factor in inducing orbital paramagnetism in this system. While such a configuration can, in principle, arise in other materials, it remains primarily associated with nodal-line semimetals and is seldom realized elsewhere.}
%Note that the DOS near $\mu=0$ is small as shown in Fig.~\ref{fig:energyshifts}.
The mechanism for this paramagnetism is discussed in detail later.
%Therefore, the present orbital paramagnetism originates from a completely new mechanism, because of the low-DOS property of the nodal line [see Fig.~\ref{fig:energyshifts}].
%
%
%We discuss the origin of it later.

%The results for $\chi_x$ at finite $\eta$ are shown 
As shown in Fig.~\ref{fig:chiz} (b),
$\chi_x$ is not an even function of the chemical potential for finite $\eta$. 
We find the relations $\chi_x(\mu, \eta)=\chi_x(-\mu, -\eta)$ and $\chi_x(\mu, \eta)=\chi_y(-\mu, \eta)$ from the analytical expressions, shown in SM \cite{sm}.
This behavior is understood as follows.
At each point on the nodal line, a Dirac dispersion $\propto \pm \sqrt{q_x^2 + q_y^2}$ is formed, 
where $q_x$-$q_y$ plane is perpendicular to the tangent of the nodal line at the point. 
Therefore, when the magnetic field is parallel to the tangent of the nodal line, 
the delta function--like orbital diamagnetism ($\propto - \delta(\mu-\varepsilon_0))$ \cite{fukuyama,koshino-ando} is negatively maximized 
with $\varepsilon_0$ being the energy of the Dirac point. 
As we can see from the symbol $\times$ in Fig.~\ref{fig:energyshifts},
when $\eta$ is positive, the energy of the nodal point at which the tangent of the nodal line 
is parallel to $\tilde{k}_x$ is negative, i.e., $\varepsilon_0<0$. 
Thus, $\chi_x$ takes the negatively maximal value at $\mu=\varepsilon_0<0$. 
The magnitude of $\chi_x$ becomes smaller as $\eta$ increases, since the region of the nodal line parallel to $\tilde{k}_x$ becomes smaller.
%is the largest when $\eta=0$ since the number of Dirac electrons near $E_F$ is the largest. 

\textit{Comparison with experiments}.---
We evaluate the orbital magnetic susceptibility for ZrSiS
using the parameters obtained by the DFT calculations.
%parameters obtained in the present and preceding 
%first-principles studies\cite{schoop2016, habe, rudenko, gudac}.
For $\chi_z$, we consider $\chi_z^{\rm tot}=\chi_z^\text{(i)}+\chi_z^\text{(ii)}+\chi_z^\text{(iii)}+\chi_{\rm Pauli}$,
where $\chi_z^\text{(X)}$ represents the contribution from the nodal line X (X=i, ii, iii),
and the Pauli paramagnetism $\chi_{\rm Pauli}=0.125 \times 10^{-4}$ emu/mol \cite{gudac} is included.
We evaluate $\chi_z^\text{(i)}$ and $\chi_z^\text{(ii)}$ from Eq.~\eqref{eq:model}.
%\footnote{\label{foot:70}\so{The temperature dependence of $\chi_z^{\rm (i)}$, $\chi_z^{\rm (ii)}$, $\chi_x^{\rm (i)}$, and $\chi_x^{\rm (ii)}$ can also be obtained from the result shown in Fig.~\ref{fig:chiz}. The susceptibility at temperature $T$ is obtained from the zero-temperature susceptibility: $\chi(\mu,T)=\int d\varepsilon (-\partial f(\varepsilon,\mu)/\partial \varepsilon)\chi(\varepsilon,T=0)$} }.
$\chi_z^\text{(iii)}$ is obtained from the model Eq.~\eqref{eq:nodal3} by the $k_z$ average of the known result for the 2D Dirac electron system \cite{raoux-piechon}
\begin{equation}
	\chi_{2D}(k_z)=\frac{e^2v_{\rm Dirac}^2}{6\pi} \frac{f(\Delta_{\rm SOI},-\varepsilon_0(k_z))-f(-\Delta_{\rm SOI},-\varepsilon_0(k_z))}{\Delta_{\rm SOI}}
	\label{eq:diracfintemp}
\end{equation}
as
$\chi_z^\text{(iii)}=\frac{V}{2\pi}\int dk_z \chi_{2D}(k_z)$ 
where $f(\varepsilon,\mu)$ is the Fermi distribution function.

\begin{figure}
    \includegraphics[width=1\linewidth]{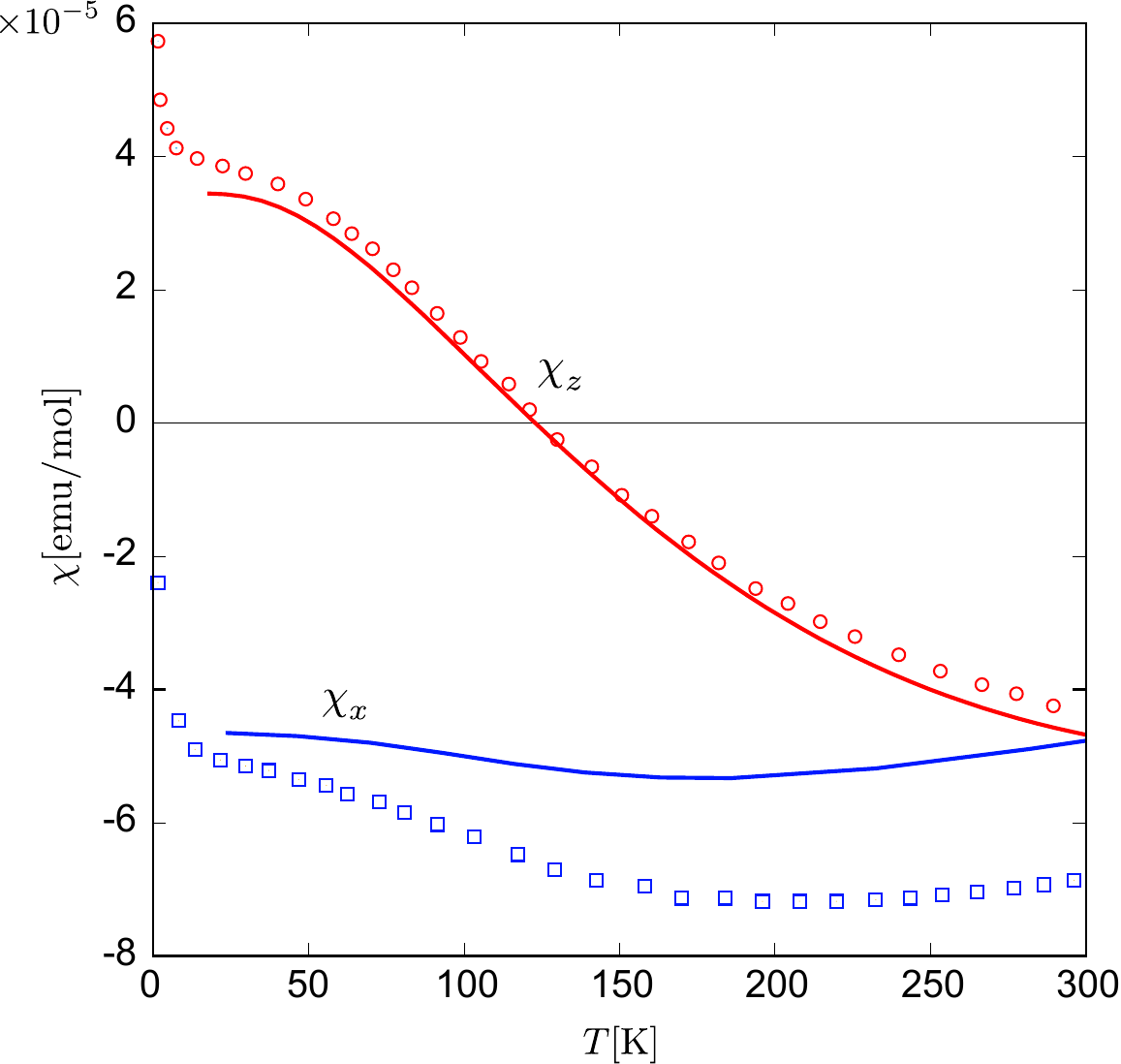}
    \caption{Temperature dependence of $\chi_z^{\rm tot}$ and $\chi_x^\text{\rm tot}$. Experimental data for $\chi_z^{\rm exp}(\bigcirc)$ and $\chi_x^{\rm exp}(\square)$ \cite{gudac} are also shown.
    \so{The enhancement of magnetic susceptibility $\chi_z$ below 10 K is owing to localized magnetic impurities \cite{gudac}, not the orbital effect.}}
    \label{fig:comparison}
\end{figure}

%In the temperature region of interest ($T<300$ K), 
%we have $k_{\rm B}T/\eta k_{\rm R}^2 \lesssim 0.1$ for the nodal line (i) which has large contribution to $\chi_z^\text{tot}$.
%Therefore, we assume that $\chi_z^\text{(i)}$ and $\chi_z^\text{(ii)}$ are constant.
%By this assumption, we determine the parameters in Eqs.~\eqref{eq:model} and \eqref{eq:nodal3}, 
%$\Delta$, $m^*$, $v_z$, and $v_{\rm Dirac}$, via 
%fitting of $\chi_z^\text{tot}$ to the experiments, shown in Fig.~\ref{fig:comparison}.
%The fitted values are shown in Table \ref{table:params} and are reasonable.
%Figure~\ref{fig:comparison} shows the temperature dependence of $\chi_z^{\rm tot}$,
%which is in good agreement with the experimental result.
%
%The results are interpreted in terms of nodal lines (i) -- (iii) as follows.
%Since we have $\tilde{\eta}^\text{(i)}=\eta^\text{(i)} k_{\rm R}^2/\Delta= 0.31$, 
%and $\tilde{\eta}^\text{(ii)}=\eta^\text{(ii)} k_{\rm R}^2/\Delta=0.06$,
%according to Fig.~\ref{fig:chiz},
% $\chi_z^\text{(i)}$ is found to be positive, 
%while $\chi_z^\text{(ii)}$ is small and negative,
%the sum of which is obtained as $\chi_{z}^{\rm (i)}+\chi_{z}^{\rm (ii)}=0.26\times 10^{-4}$ emu/mol.
From the DFT calculations \soo{[see Fig.~\ref{fig:nlgap}]}, we obtain 
%$k_{\rm R}=0.16$ \AA$^{-1}$, 
$\tilde{\eta}^\text{(i)}=0.31$ for (i), $\tilde{\eta}^\text{(ii)}=0.06$ for (ii), and
$\Delta_\text{SOI}=15$ meV and $\varepsilon_0(k_z)=(-0.025-0.07|k_z|c/\pi)$ eV for (iii), respectively.
\soo{Note that these parameters ensure that the Fermi level lies outside the Dirac mass gap along (iii).}
The other parameters are adjusted to fit the experimental data \cite{gudac}, which are 
$\Delta=0.8$ eV, $m^*/m_0=0.12$, $v_z/c_0=4\times 10^{-4}$, and $v_{\rm Dirac}/c_0=9\times 10^{-4}$ with $m_0$ and $c_0$ being the bare electron mass and the speed of light in vacuum, respectively.
%Therefore, $k_{\rm R}=0.16$ \AA$^{-1}$.
Figure~\ref{fig:comparison} shows the results of $\chi_z^\text{tot}$ and $\chi_x^\text{tot}$, \so{where $\chi_z^{\rm tot}$ agrees well with the experimental data quantitatively and $\chi_x^{\rm tot}$ captures the observed trends qualitatively.}

We now discuss each contribution in $\chi_z^\text{tot}$.
Since we have $\tilde{\eta}^\text{(i)}=0.31$ and $\tilde{\eta}^\text{(ii)}=0.06$, near $T=0$, $\chi_z^\text{(i)}$ is positive, while $\chi_z^\text{(ii)}$ is small and negative, as shown in Fig.~\ref{fig:chiz}(a).
Their sum is $\chi_z^\text{(i)}+\chi_z^\text{(ii)}=0.26\times 10^{-4}$ emu/mol, which is the main contribution for orbital paramagnetism near $T=0$.
For temperatures below 300 K, $\chi_z^\text{(i)}+\chi_z^\text{(ii)}$ is almost constant.
The temperature dependence of $\chi_z^\text{tot}$ is attributed to $\chi_z^\text{(iii)}$.
At low temperatures, 
$\chi_z^\text{(iii)}$ is negative due to Dirac electrons in the $k_x$-$k_y$ plane 
but small because the chemical potential is slightly outside the gap [see Fig.~\ref{fig:nlgap} \mo{(d)}].
%therefore, $\chi_z^\text{tot}>0$.
As the temperature increases, the diamagnetism from $\chi_z^\text{(iii)}$ grows 
because of the smearing as expressed by Eq.~\eqref{eq:diracfintemp}.
This leads to negative $\chi_z^\text{tot}$ when $T>120$ K.

\begin{comment}
\begin{table}
	\caption{Fitted parameters for Eq.~\eqref{eq:model} and Eq.~\eqref{eq:nodal3}.}
	\label{table:params}
	\begin{ruledtabular}
		\begin{tabular}{cccc}
			$\Delta$  & $m^*/m_0$ \footnotemark[1] & $v_z/c_0$ \footnotemark[2] & $v_{\rm Dirac}/c_0$ \footnotemark[2]  \\
			\colrule
			0.8 eV  &0.12 & 4$\times 10^{-4}$ & $9\times 10^{-4}$ \\
		\end{tabular}
	\end{ruledtabular}
	\footnotemark[1]{$m_0$ is the bare electron mass.}
	\footnotemark[2]{$c_0$ is the speed of light in vacuum.}
\end{table}
%According to this fitting, the orbital paramagnetism and its temperature dependence of ZrSiS are explained as follows:
\end{comment}

Note that the effect of the gap due to SOI is neglected in the model Eq.~\eqref{eq:model} used in the above calculations.
This treatment is justified, as the gap is small ($\lesssim0.03$ eV) compared with $\Delta$ ($=0.8$ eV), which limits its effect on $\chi_z$ in Fig.~\ref{fig:chiz} to the narrow range $|\mu|/\Delta\lesssim 0.04$.
\so{Explicit evaluation of the effect of gap opening is provided in SM \cite{sm}.
Even in the presence of band hybridization, the residual nodal-line features continue to play a crucial role, leaving a strong imprint on the 
interband response and still leading to paramagnetism.}
%As shown in Fig.~\ref{fig:nlgap}(b), the gap is small ($\lesssim 0.03$ eV) in the nodal line (i),
%so that its effect in $\chi_z$ shown in Fig.~\ref{fig:chiz}(a) is limited in a small range of $\mu/\Delta$ 
%(for example $|\mu|/\Delta\lesssim 0.04$).
%Therefore, we can safely neglect the effect of the small gap in the effective model of Eq.~\eqref{eq:model}.

%As for $\chi_x^{\rm tot}$, 
Next, we consider %the magnetic susceptibility in the $x$ direction defined by
$\chi_x^\text{tot}=\chi_x^\text{(i)}+\chi_x^\text{(ii)}+\chi_{\rm Pauli}$,
where we have assumed $\chi_x^\text{(iii)}=0$ since 
the magnetic field is perpendicular to \soo{the Dirac cone along the nodal line (iii)}.
We can see that the negative value of $\chi_x^\text{tot}$ in Fig.~\ref{fig:comparison} originates from $\chi_x$ \mo{of the nodal lines (i) and (ii)} shown in Fig.~\ref{fig:chiz} (b).
%\soo{
%%Note that $\chi_x^{\rm (i,ii)}$ does not exhibit as strong a chemical potential dependence as 
%%$\chi_z^{\rm (iii)}$. This difference is attributed to the energy configuration of the Dirac points: 
%Note that Dirac points along (i) and (ii) cross the Fermi level, whereas the Fermi level lies outside the gap throughout (iii) as shown in Fig.~\ref{fig:nlgap}(d).
%This difference leads to the distinct temperature dependence between $\chi_x^{\rm (i,ii)}$ and $\chi_z^{\rm (iii)}$.}
%%the nodal line (iii) forms Dirac electrons in the $k_x$-$k_y$ plane.
%%
%%Each contribution is similarly evaluated as $\chi_z$, and the result is shown in Fig.~\ref{fig:comparison}, reproducing the experiments well.
%%Thus, we consider that $\chi_x^{\rm tot}$ is basically explained by the 
%%Dirac electrons on the nodal lines (i) and (ii).
\mo{Furthermore, we find that $\chi_x^\text{(i)}+\chi_x^\text{(ii)}$ exhibits a much weaker temperature dependence than $\chi_z^\text{(iii)}$ discussed above, although in both cases the nodal lines include segments parallel to the magnetic field. 
%This difference arises from the energetic positions of the nodal lines 
The essential difference between the nodal lines (i) and (ii), on the one hand, and (iii), on the other hand, is 
the energy of the Dirac points
relative to $E_F$: (i) and (ii) cross $E_F$, whereas (iii) lies below $E_F$ [See Fig.~\ref{fig:nlgap} (c) and (d)]. 
%As a result, there is no diamagnetic contribution at zero temperature \cite{nakamuradirac,koshino-ando}, leading to a temperature dependence distinct from that of $\chi_x^\text{(i)}+\chi_x^\text{(ii)}$
%Note that the temperature dependence of (iii) is obtained by subtracting $\chi_z^\text{(i)}+\chi_z^\text{(ii)}+\chi_{\rm Pauli}(=0.385\times 10^{-4}$ emu/mol) from the result in Fig.~\ref{fig:comparison}.}
This difference leads to the difference of the temperature dependencies between 
$\chi_x^{\rm (i)} + \chi_x^{\rm (ii)}$ and $\chi_z^{\rm (iii)}$.}

%Reflecting Tateishi's comment
\begin{comment}
\so{
Finally, we discuss the agreement and discrepancy between the theory and experiments.
Although our model is based on DFT calculations, we use simplified models to capture the essence of the observed orbital paramagnetism. Nevertheless, $\chi_z^{\rm obs}$ is in good agreement with experimental results. This is because, as shown in Fig.~\ref{fig:chiz} (a), $\chi_z$ shows only a weak dependence on the chemical potential. This suggests that $\chi_z^{\rm obs}$ is robust against small variations in the omitted details. In contrast, as illustrated in Fig.~\ref{fig:chiz} (b), $\chi_x$  exhibits a relatively sharp peak, suggesting that such details can cause a modest change in $\chi_x^{\rm obs}$.
}
\end{comment}
\coauthor{
The models~\eqref{eq:model} and \eqref{eq:nodal3}, based on DFT calculations,
capture the essential features of the material.
Our calculations achieve quantitative agreement
with the experimentally measured $\chi_z$ and successfully reproduce the characteristic behavior of
$\chi_x$. These results support the validity of the circular approximation for
the nodal-line geometry and confirm that the energy variation and negative 
curvature in the dispersion are key ingredients for the observed orbital paramagnetism.
We note that this quantitative agreement can be attributed to \soo{the weak dependence 
of $\chi_z^{\rm (i)}$ and $\chi_z^{\rm (ii)}$} on the chemical potential $\mu$, 
as shown in Fig.~\ref{fig:chiz}(a), which leads to robustness 
against small variations in the omitted details.
In contrast, as illustrated in Fig.~\ref{fig:chiz}(b), \soo{$\chi_x^{\rm (i)}$ and $\chi_x^{\rm (ii)}$} exhibit
stronger $\mu$ dependence, suggesting that \coauthormatsu{more realistic models} \soo{for nodal line (i) and (ii)} would be 
necessary to reproduce the experimental results with higher precision.
}

\begin{figure}
	\includegraphics[width=\linewidth]{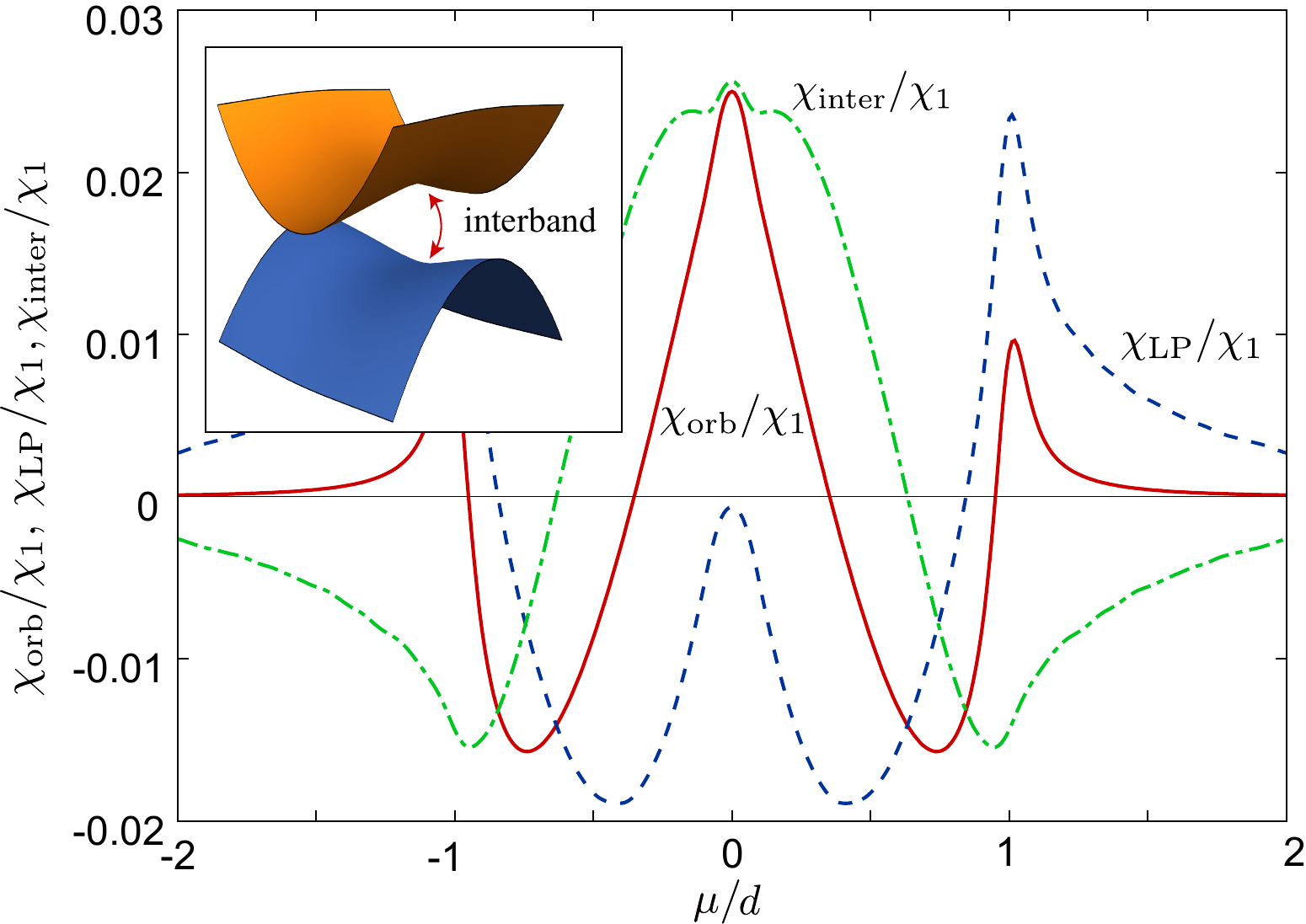}
	\caption{Total orbital magnetic susceptibility $\chi_{\rm orb}$, LP contribution 
	$\chi_{\rm LP}$, and the interband contribution $\chi_{\rm inter}\equiv \chi_{\rm orb}-\chi_{\rm LP}$ 
    of the saddle point model Eq.~\eqref{eq:interbandmodel} 
	in units of $\chi_1=e^2/\hbar^2$ with $k_{\rm B}T=1/40$,
	where we have used the unit of $\alpha=d=1$.
    Numerical integration has been conducted on a region $|k_x|\leq k_c$ and $|k_y|\leq k_c$
    with $k_c=4$, showing weak cutoff dependence.
	Inset: Energy dispersions for Eq.~\eqref{eq:interbandmodel}.
	}
    \label{fig:saddlepoint}
\end{figure}

\textit{Effective model and discussions.---} To understand the mechanism underlying the interband orbital paramagnetism,
which is essential for $\chi_z$ in ZrSiS,
we introduce an effective model,
\begin{equation}
	H=\alpha \frac{k_x^2 - k_y^2}{2}\sigma_0 + \alpha \frac{k_x^2 + k_y^2}{2} \sigma_z + d \sigma_x,
	\label{eq:interbandmodel}
\end{equation}
where $\alpha$ and $d$ represent the energy scale and band hybridization,
respectively.
This model is obtained from a small $(k_x,k_y)$ expansion of Eq.~\eqref{eq:model} 
around $(k_x, k_y, k_z)=(\pm k_{\rm R}, 0, \so{k_{z0}})$, for some fixed $k_{z0}\neq 0$.
%\textcolor{green}{As shown in the inset of Fig.~\ref{fig:saddlepoint}, both the conduction and valence bands exhibit saddle points that differ slightly in number from those in the original model Eq.~\eqref{eq:model}, while preserving the characteristic band structure.}
%Using the Fukuyama formula Eq.~\eqref{eq:fukuyama}, 
Figure~\ref{fig:saddlepoint} shows the orbital magnetic susceptibility in the $z$ direction, $\chi_{\rm orb}$,
calculated by Eq.~\eqref{eq:fukuyama}.
%is obtained in the present model,
%which is shown in Fig.~\ref{fig:saddlepoint}.
%The intraband contribution obtained by the 
The LP (intraband) contribution $\chi_{\rm LP}$ \cite{landau,peierls, raoux-piechon, ogata-fukuyama} 
and the interband contribution ($\chi_{\rm inter}=\chi_{\rm orb}-\chi_{\rm LP}$) are
also shown in Fig.~\ref{fig:saddlepoint}.
Our results show that this model exhibits orbital paramagnetism near $\mu/d\sim 0$ where the ground state is insulating,
which is not explained by the intraband effect.
The dependence on the chemical potential resembles that observed in Fig.~\ref{fig:chiz}(a), clearly showing that the orbital paramagnetism in Fig.~\ref{fig:chiz}(a) originates from the interband effect between the saddle points.
\so{This new mechanism stands in stark contrast to orbital paramagnetism arising from the Van Hove singularity in single-band 2D systems \cite{vignaleparasaddlepoint,raoux-piechon} or flatband systems \cite{Rhim2020,piechon},
%in which the Van Hove singularity leads to the orbital paramagnetism 
as it does not rely on divergent DOS or even the presence of Fermi surfaces.} 
%Therefore, the present result of the interband orbital paramagnetism demonstrates the new mechanism 
%that has not been known before.
\so{The differences between these mechanisms of orbital paramagnetism are summarized in Table~\ref{table:difference}.}
\begin{comment}
\begin{table}
	\caption{Difference of mechanisms of orbital paramagnetism}
	\label{table:difference}
	\begin{ruledtabular}
		\begin{tabular}{llll}
		  Mechanism & Van Hove Singularity& Flatband & Interband negative  \\
                &\cite{vignaleparasaddlepoint} & \cite{piechon} & curvature (present) \\
			\colrule
			Inter- or & Intraband & Interband & Interband \\
            intraband & & & \\
            Fermi surface & Necessary & Necessary & Unnecessary \\
            DOS & Divergent & Divergent & Non-divergent
		\end{tabular}
	\end{ruledtabular}
\end{table}
\end{comment}

\begin{table}
	\caption{Comparison of orbital paramagnetism mechanisms across different systems}
	\label{table:difference}
	\begin{ruledtabular}
		\begin{tabular}{lccc}
		  System & Type& Fermi surface & DOS  \\
			\colrule
			\makecell[l]{Van Hove \\ singularity \cite{vignaleparasaddlepoint}} & Intraband & Necessary & Divergent \\
            Flatband \cite{piechon} & Interband & Necessary & Divergent\\
            Nodal line (present) & Interband & Not required & Non-divergent \\
		\end{tabular}
	\end{ruledtabular}
\end{table}

%On the other hand, in the case of ZrSiS, the enhancement of DOS is absent.
%Nevertheless, the orbital paramagnetism is observed, which is attributed to 
%the \textit{interband} effect between two bands with negative curvature.

\textit{Summary}.---
We have studied the orbital magnetism in ZrSiS 
based on the DFT calculations and an effective model.
Our results elucidate three anomalies observed in the orbital magnetism,
the large orbital paramagnetism without DOS enhancement, temperature dependence, and paramagnetic-to-diamagnetic anisotropy.
We have found that the orbital paramagnetism in the $C_4$ axial direction ($k_z$ direction) at low temperatures
is attributed to an interband effect. 
This mechanism is novel in that the orbital paramagnetism is not accompanied by an enhancement of DOS.
%and thus of Pauli paramagnetism
%which usually obscures the orbital contribution.
%In contrast, our mechanism allows clear observation of orbital paramagnetism. 
\coauthor{
Therefore, the Pauli paramagnetism is suppressed, enabling the clear observation
of orbital paramagnetism.
This interband effect is understood in terms of a simple effective two-band model. 
In this model, the negative curvature, which arises from variation of the energy of the node points along the nodal line in the material-specific model, induces the anomalous orbital response.
This variation, which persists even when a small band hybridization is present, is typical of nodal-line semimetals and difficult to achieve in conventional systems.
Orbital paramagnetism is one of the most prominent features of nodal-line semimetals,
and importantly, it is experimentally accessible.}
%\textcolor{red}{regardless of the small hybridization gap}.
%This structure is ubiquitous in realistic nodal-line materials.
%Therefore, these materials are promising platform where we can observe orbital paramagnetism.
%Orbital paramagnetism is one of the experimentally observable prominent features of nodal-line semimetals.

%%%%%%%%%%%%%%%%%%%%%%%%%%%%
\begin{acknowledgments}
We are grateful to Y.\ Suzumura, M.\ Hayashi, 
Z.\ Hiroi, T.\ Osada, N.\ Tsuji, 
N.\ Kawashima,
and H.\ Shinaoka for their insightful discussions. 
This work was supported by Grants-in-Aid for Scientific Research 
from the Japan Society for the Promotion of Science (No.\ JP22J15355, No.\ JP18H01162, No.\ JP21H05191, No.\ JP18K03482, No.\ JP17H02912, No.\ JP21H01003, No.\ JP22K03447,  and No.\ JP23H04869).
\end{acknowledgments}
\bibliography{refv2}
\end{document}

\title{Supplemental Material for ``Orbital Paramagnetism without Density of State Enhancement in Nodal-line Semimetal ZrSiS''}

\author{Soshun Ozaki}
\affiliation{Department of Basic Science, University of Tokyo, Meguro, Tokyo 153-0041, Japan}
\author{Hiroyasu Matsuura}
\affiliation{Department of Physics, University of Tokyo, Bunkyo, Tokyo 113-0033, Japan}
\author{Ikuma Tateishi}
\affiliation{Department of Physics, Osaka University, Toyonaka, Osaka 560-0043}
\author{Takashi Koretsune}
\affiliation{Department of Physics, Tohoku University, Sendai, Miyagi 980-8578}
\author{Masao Ogata}
\affiliation{Department of Physics, University of Tokyo, Bunkyo, Tokyo 113-0033, Japan}
\affiliation{Trans-Scale Quantum Science Institute, University of Tokyo, Bunkyo, Tokyo 113-0033, Japan}

\maketitle
\section{Details of the first-principles calculation}
\begin{figure}
    \includegraphics[width=\linewidth]{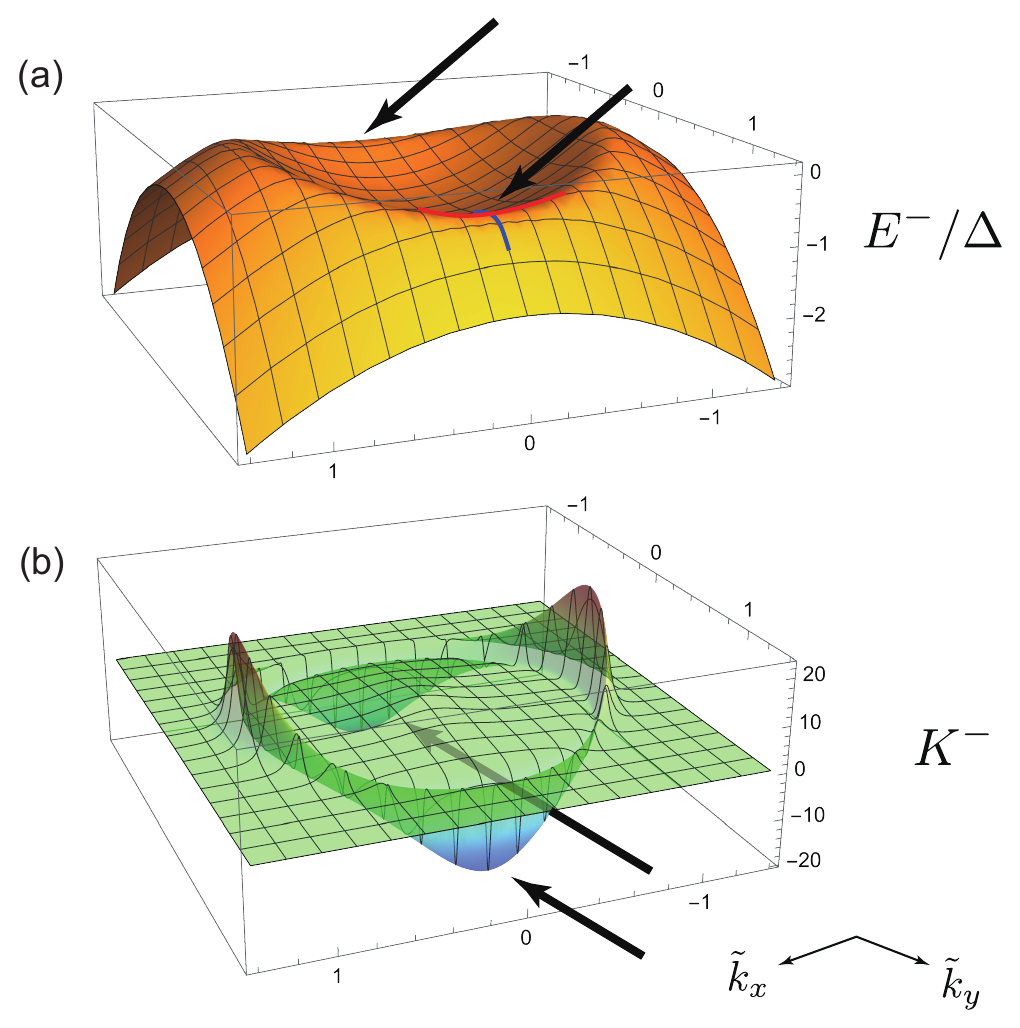}
    \caption{(a) Energy dispersion for lower band in units of $\Delta$ at $\tilde{k}_z=0.2$ and $\tilde{\eta}=0.3$. Two saddle points, $(\tilde{k}_x, \tilde{k}_y)=(\pm 1,0)$ are shown with arrows.
    At $(\tilde{k}_x,\tilde{k}_y)=(0,1)$, the upward (downward) convex contour is indicated by a red (blue) curve.
    (b) Curvature $K^-$ of the dispersion (a). Negative curvature is also indicated with arrows.}
    \label{fig:gausscurvature}
\end{figure}

%Our calculation is performed by QUANTUM ESPRESSO \cite{qe}, which uses the density functional theory
%\cite{hohenbergkohn, kohnsham}
%with spin-orbit interaction (SOI).
%For the exchange-correlation term, the generalized gradient approximation ... (correct?).
%The Kohn-Sham orbitals are expanded with plane waves and the cutoff energies are ?? for wave functions and
%charge density, respectively.
%The ${\bm k}$-point grid on the Brillouin zone is taken as ??.
The density functional theory (DFT) calculations are performed using the \textsc{Quantum-ESPRESSO} package\cite{Giannozzi2017}.
For the experimental crystal structure with lattice parameters of $a=3.55$ \AA\ and $c=8.07$ \AA\cite{Onken1964},
the electronic structures with spin-orbit couplings are obtained within the generalized gradient approximation with Perdew-Burke-Ernzerhof exchange-correlation functionals\cite{pbe1996} and ultrasoft pseudopotentials\cite{vanderbilt1990} in the PSLibrary\cite{Corso2014}.
The cutoff energies for wave functions and charge density are 50 Ry and 500 Ry, respectively.
The ${\bm k}$-point grid in the Brillouin zone is set to $12\times 12\times 6$.
To calculate the nodal lines, the tight-binding model obtained using the \textsc{Wannier90} package\cite{Pizzi2020,koretsune2023} is employed.
%
%
%%%%%%%%%%%%%%%%%%
\so{
\section{Negative curvature of energy dispersion}
In this section, we demonstrate the existence of negative curvature in the vicinity of the nodal line
in model Eq.~(1) in the main text.
\soo{Under a magnetic field along the $z$ direction, only the in-plane ($k_x$-$k_y$) orbital 
motion contributes since the $k_z$-dependent part is decoupled.
It is therefore natural to define the Gauss curvature in the $E$-$k_x$-$k_y$ space, i.e., on a two-dimensional slice at fixed $k_z$. The Gauss curvature is given by}
\begin{equation}
    K^{\pm}=\frac{\varepsilon^\pm_{xx}\varepsilon^\pm_{yy}-(\varepsilon^\pm_{xy})^2}{(1+(\varepsilon^\pm_x)^2+(\varepsilon^\pm_y)^2)^2},
\end{equation}
where $\varepsilon^\pm=E^\pm/\Delta$, $E^\pm$ is the energy dispersion for conduction/valence band, and we have used abbreviations, $\varepsilon^\pm_{x}=\frac{\partial \varepsilon^\pm}{\partial \tilde{k}_x}$, $\varepsilon^\pm_y=\frac{\partial \varepsilon^\pm}{\partial \tilde{k}_y}$, $\varepsilon^\pm_{xx}=\frac{\partial^2 \varepsilon^\pm}{\partial \tilde{k}_x^2}$, $\varepsilon^\pm_{yy}=\frac{\partial^2 \varepsilon^\pm}{\partial \tilde{k}_x^2}$, and $\varepsilon^\pm_{xy}=\frac{\partial^2 \varepsilon^\pm}{\partial \tilde{k}_x \partial \tilde{k}_y}$.
The normalized energy dispersion and curvature of the valence band are shown in Fig.~\ref{fig:gausscurvature}.
These results show that there exist four stationary points in the $k_x$-$k_y$ plane: 
$(\pm k_{\rm R},0)$, among which $(0,\pm k_{\rm R})$ are saddle points,
exhibiting negative curvature.}

%Since the numerator of Eq.~\eqref{} is contained in the Landau-Peierls formula,  
%the intraband orbital magnetic susceptibility is subject to the sign of this curvature.
%For interband orbital magnetism 

\section{Analytical results for orbital magnetic susceptibility}
We present the detailed analytical calculations of $\chi_z$ and $\chi_x$.
First, we consider a general form of the model Hamiltonian,
\begin{equation}
	{\mathcal H} = (a k_x^2 + b k_y^2-\Delta) \sigma_{z} + \hbar v_z k_z \sigma_{x} + \eta (k_x^2 - k_y^2) \sigma_0
,  \label{eq:model}
\end{equation}
After calculating the trace and Matsubara summation, we obtain
\begin{equation}
  \chi_z = \frac{1}{2\pi^2}\frac{e^2}{\hbar^2}\frac{\sqrt{ab}}{\hbar v_z} \int_{-\Lambda}^{\Lambda} dp_z 
	\sum_{\pm} \left(\pm \frac{\Delta}{6\epsilon_z}\right)f(\pm \epsilon_z). 
\end{equation}
\begin{align}
	\chi_x  
	=&\frac{e^2}{\hbar^2}\frac{\hbar v_z}{24\pi^2}\sqrt{\frac{b}{a}} 
\biggr\{-2 + \sum_\pm \int_{-\Lambda}^{\Lambda}dp_z  \biggr[ \mp \frac{\epsilon_z}{\Delta^2}f(\pm \epsilon_z) \nonumber \\
	&\pm 3\Delta \int_{\Delta}^{\infty} dx 
	\frac{\sqrt{x^2+p_z^2}}{x^4}f(\pm \sqrt{x^2+p_z^2}) \biggr] \biggr\} \nonumber \\
	&+\frac{e^2}{\hbar^2} \frac{\hbar v_z}{3\pi}\sqrt{\frac{b}{a}} f'(0),
\end{align}
and
\begin{align} 
	\chi_y=(a/b)\chi_x,
\end{align}
where $p_z=\hbar v_z k_z$, $\epsilon_z=\sqrt{\Delta^2+p_z^2}$, and $\Lambda$ is the cutoff of $p_z$ integral
determined by the band width.
The functions $f(x)$ and $f'(x)$ are the Fermi distribution function and its derivative, respectively.

In particular at $T=0$, the orbital susceptibility is further calculated, and we obtain
\begin{widetext}
\begin{equation}
\chi_z =
	\begin{dcases}
		-\frac{1}{6\pi^2}\frac{e^2}{\hbar^2}\frac{\sqrt{ab} \Delta}{\hbar v_z} 
		\ln{\frac{\Lambda + \sqrt{\Lambda^2+\Delta^2}}{\Delta}} & {\rm for}\,  |\mu| \le \Delta \\
		-\frac{1}{6\pi^2}\frac{e^2}{\hbar^2}\frac{\sqrt{ab} \Delta}{\hbar v_z} 
		\ln{\frac{\Lambda + \sqrt{\Lambda^2+\Delta^2}}{|\mu| + \sqrt{\mu^2 - \Delta^2}}} 
		& {\rm for}\, \Delta \le |\mu| \le \sqrt{\Delta^2+\Lambda^2} %\\
		%0 & {\rm for}\, \sqrt{\Delta^2+\Lambda^2}\le |\mu|,
	\end{dcases}
\label{eq:smchiz}
\end{equation}
and
\begin{eqnarray}
\chi_x =
	\begin{dcases}
-\frac{e^2}{\hbar^2}\frac{\hbar v_z}{2\pi^2}\sqrt{\frac{b}{a}} \frac{2\pi \Delta}{3} \delta(\mu) 
		+\frac{e^2}{\hbar^2}\frac{\hbar v_z}{24\pi^2} \sqrt{\frac{b}{a}} f_1(\Delta) & {\rm for}\, |\mu| \le \Delta \\
		\frac{e^2}{\hbar^2}\frac{\hbar v_z}{24\pi^2} \sqrt{\frac{b}{a}} f_2(\mu, \Delta) & {\rm for}\, \Delta \le |\mu| \le \sqrt{\Delta^2+\Lambda^2} \\
		%- \frac{e^2}{\hbar^2}\frac{\hbar v_z}{24\pi^2} \sqrt{\frac{b}{a}} \frac{2\Delta}{\Lambda}
		%\left(1+\frac{2\Lambda^2 -\mu^2}{|\mu|\sqrt{\mu^2-\Lambda^2}} \right) & {\rm for} \, \sqrt{\Delta^2+\Lambda^2}\le |\mu|,
	\end{dcases} \label{eq:smchix}
\end{eqnarray} 
where we have defined 
\begin{equation}
    %wrong 2024/4/10   f_1(\Delta)= \frac{2}{\Lambda}(\sqrt{\Lambda^2+\Delta^2}-\Delta-\Lambda)- 2\ln \frac{\Lambda+\sqrt{\Lambda^2+\Delta^2}}{\Delta}
    f_1(\Delta)= \frac{2}{\Lambda}(\sqrt{\Lambda^2+\Delta^2}-\Delta-\Lambda)- 2\ln \frac{\Lambda+\sqrt{\Lambda^2+\Delta^2}}{\Delta}, %Lambda is not wrong
\end{equation}
and
\begin{equation}
    %wrong 2024/4/10  f_2(\mu, \Delta)= - \frac{4\sqrt{\mu^2-\Delta^2}}{|\mu|} - 2\ln \frac{\Lambda+\sqrt{\Lambda^2+\Delta^2}}{\sqrt{\mu^2-\Delta^2}+|\mu|}
    %       +\frac{2}{\Lambda}(\sqrt{\Lambda^2+\Delta^2}-\Delta - \Lambda)
    f_2(\mu, \Delta)= - \frac{4\sqrt{\mu^2-\Delta^2}}{|\mu|} - 2\ln \frac{\Lambda+\sqrt{\Lambda^2+\Delta^2}}{\sqrt{\mu^2-\Delta^2}+|\mu|}
    +\frac{2}{\Lambda}(\sqrt{\Lambda^2+\Delta^2}-\Delta -\Lambda).
\end{equation} 
We can see that Eqs.~\eqref{eq:smchiz} and \eqref{eq:smchix} reproduce the previous study \cite{tateishi}
in the limit of $\Lambda_R\to \infty$ and $\Lambda_z \to \infty$.

Next, we consider the case for finite $\eta$, which is used for the numerical evaluation.
In the following, we set $a=b$ for simplicity.
After calculating the trace, we obtain
\begin{align}
    \chi_z =& \frac{e^2}{\hbar^2} k_BT \sum_{n {\bm k}}  \frac{32 k_x^2 k_y^2}{D^2}
    \left[(a^2+\eta^2)^2 + 4a^2\eta^2 + \frac{2h_1({\bm k}, i\epsilon_n)}{D}(a^2+\eta^2)
    + \frac{[h_1({\bm k}, i\epsilon_n)]^2}{2D^2}\right]  \label{eq:chizweta}\\
    \chi_x =& \frac{e^2}{\hbar^2} k_B T \sum_{n {\bm k}} \frac{8k_y^2}{D^2} \hbar^2 v_z^2
    \left(-a^2+\eta^2 + \frac{2\hbar^2 v_z^2 k_z^2  h_2({\bm k}, i\epsilon_n)}{D^2} \right) \label{eq:chixweta}\\
    \chi_y =& \frac{e^2}{\hbar^2} k_B T \sum_{n {\bm k} }\frac{8k_x^2}{D^2} \hbar^2 v_z^2
    \left(-a^2+\eta^2 + \frac{2\hbar^2 v_z^2 k_z^2  h_3({\bm k}, i\epsilon_n)}{D^2} \right), \label{eq:chiyweta}
\end{align}
where $D = [i\epsilon_n + \mu- \eta(k_x^2 - k_y^2)]^2- (ak_x^2+bk_y^2-\Delta)^2 - \hbar^2 v_z^2 k_z^2$ and 
\begin{eqnarray}
    h_1 ({\bm k}, i\epsilon_n)&=& 4\{ a^2 (ak_x^2 + ak_y^2 -\Delta)^2 - \eta^2[i\epsilon_n + \mu -\eta(k_x^2 - k_y^2)]^2 \} \\
    h_2 ({\bm k}, i\epsilon_n)&=& 4\{ a (ak_x^2 + ak_y^2 -\Delta)     - \eta  [i\epsilon_n + \mu -\eta(k_x^2 - k_y^2)]  \}^2 \\
    h_3({\bm k}, i\epsilon_n)&=& 4\{ a (ak_x^2 + ak_y^2 -\Delta)     + \eta  [i\epsilon_n + \mu -\eta(k_x^2 - k_y^2)]  \}^2,
\end{eqnarray}
\end{widetext}
and $\chi_y$ is obtained by substituting $x\leftrightarrow y$ and $\eta \to -\eta$ into $\chi_x$.
By the numerical evaluation, we obtain the results in Figs.~3 and 4 in the main text.

%%%%%%%%%%%%%%%%%%%%%%%%%%
\section{Numerical integration}
For numerical integration for the formula Eq.~(3) in the main text, we rewrite quantities in dimensionless forms. 
For this purpose, we define the dimensionless Hamiltonian
\begin{align}
	\frac{H}{\Delta} &= \left[\frac{\hbar^2}{2m^* \Delta}(k_x^2+k_y^2)-1 \right] \sigma_z + \frac{\hbar v_z}{\Delta} k_z \sigma_x
	+\frac{\eta}{\Delta}(k_x^2-k_y^2) \nonumber \\
	&=(\tilde{k}_x^2 + \tilde{k}_y^2 - 1)\sigma_z + \tilde{k}_z \sigma_x + \tilde{\eta}(\tilde{k}_x^2-\tilde{k}_y^2),
\end{align}
where we have defined
\begin{equation}
	\tilde{k}_x = k_x l_0, \quad \tilde{k}_y= k_y l_0, \quad \tilde{k}_z=k_z l_z, \quad \tilde{\eta}=\eta k_{\rm R}^2/\Delta
\end{equation}
with
\begin{equation}
	l_0 = \frac{\hbar}{\sqrt{2m^*\Delta}}=\frac{1}{k_{\rm R}}, \quad l_z = \frac{\hbar v_z}{\Delta}.
\end{equation}
Furthermore, we define dimensionless Green's function and current operators for the Fukuyama formula,
\begin{align}
	&\tilde{\mathcal{G}}=\Delta \mathcal{G}, \quad \tilde{\gamma_x}=(l_0 \Delta)^{-1}\gamma_x, \nonumber \\
	&\tilde{\gamma_y}=(l_0 \Delta)^{-1}\gamma_y, \quad \tilde{\gamma_z}=(l_z \Delta)^{-1}\gamma_z.
\end{align}
Then, we obtain $\chi_x$ and $\chi_z$ in the dimensionless forms as
\begin{align}
	\chi_x &= \chi_0 \frac{k_B T}{\Delta} \sum_n \frac{1}{(2\pi)^3} 
	\int d \tilde{\bm k} {\rm Tr} (\tilde{\mathcal{G}}\tilde{\gamma_y}\tilde{\mathcal{G}}\tilde{\gamma_z})^2, \\
	\chi_z &= \chi_0  \left(\frac{1}{k_{\rm R} l_z}\right)^2 \frac{k_B T}{\Delta} \sum_n \frac{1}{(2\pi)^3} 
	\int d \tilde{\bm k} {\rm Tr} (\tilde{\mathcal{G}}\tilde{\gamma_x}\tilde{\mathcal{G}}\tilde{\gamma_y})^2,
\end{align}
where we have set the unit of magnetic susceptibility as $\chi_0=\frac{e^2}{\hbar^2}Vl_z\Delta=0.912\times 10^{-4}$ emu/mol with $V=3.07 \times 10^{25}$ \AA$^3$/mol,
$l_0=6.3$ \AA, and $l_z=1.0$ \AA.  
We employ cylindrical coordinates and set the cutoff for ${\bm k}$ integral to $\Lambda_{\rm R}=1000k_{\rm R}$ 
for the radial direction and $\Lambda_z=(l_z)^{-1}$ in the $k_z$ direction. 

\so{
\section{effect of SOI on $\chi_z$}
In this section, we discuss the gap due to SOI on the orbital magnetic susceptibility $\chi_z$.
The Hamiltonian with SOI gap reads
\begin{align}
    {\mathcal H}=& \biggl[\frac{\hbar^2}{2m^*}(k_x^2 + k_y^2)-\Delta\biggr] \sigma_{z} + \hbar v_z k_z \sigma_{x} + \eta (k_x^2 - k_y^2) \sigma_0 \nonumber \\
    &+ \Delta_{\rm SOI}\sigma_y.
\end{align}
Using the same numerical method as in the main text, we calculate $\chi_z$.
Figure~\ref{fig:chiz_soi} shows the results obtained.
We find that the effect of the SOI gap on $\chi_z$ is not significant, particularly when $\eta\simeq 0$, 
which justifies our assumption.
\begin{figure}[H]
    \includegraphics[width=\linewidth]{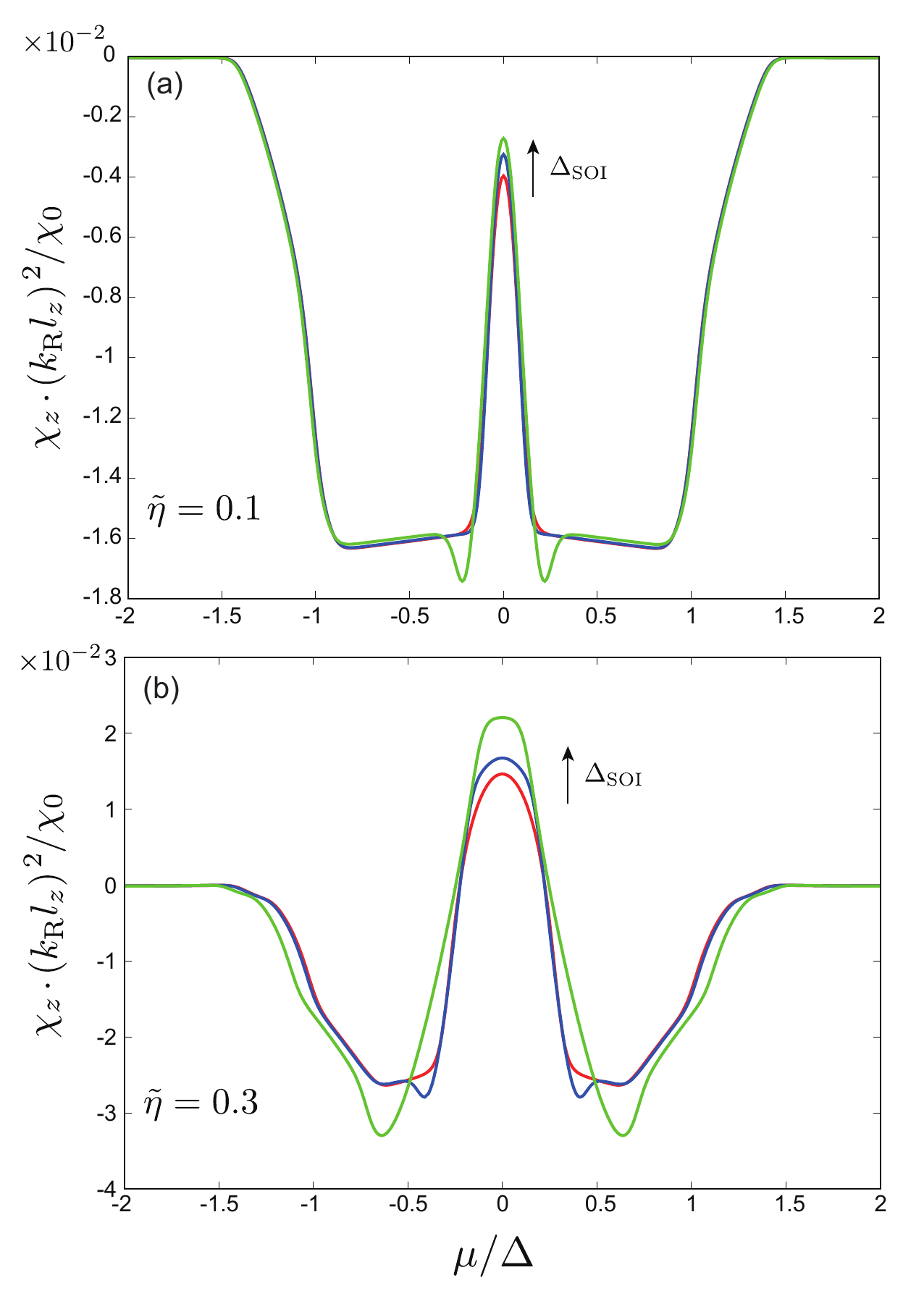}
    \caption{Orbital magnetic susceptibility $\chi_z$ in the presence of a SOI gap for the cases of (a) $\tilde{\eta}=0.1$ and (b) $\tilde{\eta}=0.3$. The red, blue, and green lines represent different sizes of the SOI gap, $\Delta_{\rm SOI}=0, 0.5 \tilde{\eta}\Delta,$ and $1.5\tilde{\eta}\Delta$, respectively, for each $\tilde{\eta}$.}
    \label{fig:chiz_soi}
\end{figure}
}

\section{gap size-scaling of orbital paramagnetism in the effective model}
\so{In this section, we discuss the scaling of $\chi_{\rm orb}$ obtained from Eq.~(5) in the main text with respect to the size of gap, $d$.
Using Eq.~(3) in the main text, we obtain the scaling $\chi_{\rm orb}(T=0,\mu=0,\lambda d)=\chi_{\rm orb}(T=0, \mu=0, d)$ for a positive parameter $\lambda>0$, different from the scaling relation for the two-dimensional Dirac electrons, $\chi_{2D}(T=0,\mu=0,\lambda d)=\lambda^{-1}\chi_{2D}(T=0,\mu=0, d)$.
This result implies that $\chi_{\rm orb}$ does not exhibit significant dependence on the gap size $d$.}

\bibliography{refv2}